\documentclass[pra,twocolumn]{revtex4}
\usepackage{amsmath,amssymb,mathrsfs}
\usepackage[dvips]{graphicx}
\usepackage{hyperref}

\newcommand{\ua}{\uparrow}
\newcommand{\da}{\downarrow}

\def\be{\begin{equation}}       \def\ee{\end{equation}}
\def\bea{\begin{eqnarray}}      \def\eea{\end{eqnarray}}
\def\nn{\nonumber}

\begin{document}

\title{Theory of quasi-one dimensional imbalanced Fermi gases}

\author{Erhai Zhao and W. Vincent Liu}
\affiliation{Department of Physics and Astronomy, University of
Pittsburgh, Pittsburgh, Pennsylvania 15260, USA}

\begin{abstract}

We present a theory for a lattice array of weakly coupled
one-dimensional ultracold attractive Fermi gases (1D `tubes') with
spin imbalance, where strong intratube quantum fluctuations invalidate
mean field theory.  We first construct an effective field theory,
which treats spin-charge mixing exactly, based on the Bethe ansatz
solution of the 1D single tube problem. We show that the 1D
Fulde-Ferrel-Larkin-Ovchinnikov (FFLO) state is a two-component
Luttinger liquid, and its elementary excitations are fractional states
carrying both charge and spin.  We analyze the instability of the 1D
FFLO state against inter-tube tunneling by renormalization group
analysis, and find that it flows into either a polarized Fermi liquid
or a FFLO superfluid, depending on the magnitude of interaction
strength and spin imbalance.  We obtain the phase diagram of the
quasi-1D system and further determine the scaling of the superfluid
transition temperature with intertube coupling.
\end{abstract}

\maketitle

\section{Introduction}
The last few years witnessed an explosion of studies in population
imbalanced (`polarized') Fermi gases, partly fueled by the MIT and
Rice experiments with resonantly interacting ultracold $^6$Li atoms
\cite{rice1,mit1}.  One of the motivations of these and following
experiments \cite{mit2,shin,rice2,mit3} is to search for the
Fulde-Ferrel-Larkin-Ovchinnikov (FFLO) state, a spatially modulated
superfluid \cite{review}. So far, no evidence of FFLO was found in
three dimensional (3D) traps. This is in agreement with the mean field
theory prediction that FFLO state only occurs in a tiny region in the
phase diagram of 3D polarized Fermi gas \cite{sheehy}.

Contrary to 3D, in one dimension (1D), Bethe ansatz \cite{orso,hu,guan}
and Density Matrix Renormalization Group (DMRG) studies
\cite{feiguin,fazio} show that a FFLO-like phase (which only has
algebraic order) occupies a rather large region in the phase diagram
of polarized attractive Fermi gas.  Taking these results together,
recently Parish {\it et al} \cite{parish} proposed that the most
promising regime to realize FFLO state is in quasi-one dimension,
i.e., a 2D lattice array of cold atomic ``tubes'' \cite{esslinger}
where the transverse tunneling $t_\perp$ is weak. They obtained the
phase diagram of the quasi-1D system using mean field theory
\cite{parish}. These works prompted the experimental search for the
FFLO phase in quasi-1D atomic gas systems.  These highly tunable
systems are also ideal for investigating the 1D-to-3D dimensional
crossover of superfluids \cite{schulz}.

A few key questions remain open regarding the quantum phases of quasi-1D atomic systems.
1) What is the phase diagram in the limit of small $t_\perp$?
In this regime, intratube quantum fluctuations is
dominantly strong \cite{schulz} so the mean field approach of \cite{parish} ceases to apply.
To find the solution, one has to first answer a related question:
2) What is the low energy effective description for the 1D FFLO phase?
An effective theory can yield analytical expressions for the zero and finite
temperature correlation functions and serve as a starting point to analyze the
instability of 1D FFLO against $t_\perp$.
Note that a field theory for 1D FFLO state was previously proposed by Yang to
describe 1D superconductors in magnetic field \cite{kyang}. In Yang's model,
spin and charge degrees of freedom are assumed decoupled.
However, we shall show that this model does not apply to attractive
Fermi gases with {\it finite} spin imbalance (`magnetization') because the
assumption of spin-charge separation breaks down \cite{fazio}.

In this paper, we answer these open questions using a combination of
several well established theoretical techniques. We first develop an effective field
theory for the 1D FFLO state, which properly treats the spin-charge mixing
in trapped two component atomic gases and fully takes into account the
intratube quantum fluctuations. Then we treat the transverse tunneling perturbatively \cite{schulz}
to obtain the phase diagram of a 2D lattice array of weakly coupled 1D
Fermi gases, which is an anisotropic 3D system.
We also derive the scaling formula for the transition temperature
of the quasi-1D FFLO state.

Our approach, based on exact solutions, is complementary to the mean
field method of Ref \cite{parish}, which applies to the opposite limit of large
$t_\perp$. Taken together, they can provide a more complete
understanding of the dimensional crossover of quasi-1D attractive
Fermi gases with finite population imbalance, and provide
guidance for the experimental search of the quasi-1D FFLO phase.

\section{Model}
We consider a gas of two-component fermionic atoms, e.g., $^6$Li atoms in
hyperfine states $|F=1/2,m_f=\pm 1/2\rangle$ (labeled with spin up and down
respectively), confined in a highly elongated
one dimensional (1D) trap interacting via an attractive contact potential,
\be
H_{GY}=-\frac{\hbar^2}{2m}\sum_{i=1}^N \frac{\partial^2}{\partial x_i^2}
+ g\sum_{i=1}^{N_\uparrow} \sum_{j=1}^{N_\downarrow}
\delta(x_i-x_j). \label{gaudin-yang}
\ee
This Hamiltonian is known as the Gaudin-Yang model \cite{cny,gaudin}.
Here $m$ is the fermion mass, and the attractive interaction stength
$g\equiv -2\hbar^2/ma_{1D}<0$, $a_{1D}$ is the effective 1D
scattering length and related to the 3D $s$-wave scattering length
$a_s$ by \cite{olsh,olsh2}
\[
a_{1D}=(1.03-\frac{a_\perp}{a_s})a_\perp,
\]
where $a_\perp=\sqrt{\hbar/m\omega_\perp}$ is the oscillator length of
the transverse
confinement of frequency $\omega_\perp\gg \omega_x$. { Let $N_\uparrow$ and
$N_\downarrow$ be the number of spin up and down particles
respectively, and $N=N_\uparrow+N_\downarrow$ the total particle number.}
We define the linear particle density $n_L=N/L$, the population imbalance
$p=(N_{\uparrow}-N_\downarrow)/N$, and the dimensionless interaction parameter
$\gamma=2/(n_La_{1D})$. These are experimentally controllable
parameters. Some of these numbers are listed in
Ref. \onlinecite{ceperley}.
{ In the above model Hamiltonian,
we have neglected the slowly varying Gaussian trap potential
in $x$ direction, $V(x)=m\omega_x^2x^2/2$, and assumed periodic
boundary condition for the system of length $L$, as we
are interested in the collective behaviors of the population imbalanced gas in
the thermodynamic limit.
The effect of the trap can be subsequently taken into account using
local density approximation.}

For technical convenience, we also consider a closely related lattice model,
the 1D attractive Hubbard model,
\be
H_U = -t\sum_{i,\sigma}(c^\dagger_{i,\sigma}c_{i+1,\sigma}+h.c.) - U\sum_i n_{i,\ua}n_{i,\da}, \label{hubbard}
\ee
with linear particle density $n_L=n/a$ and population imbalance $p$. Here,
$n=n_{\ua}+n_\da$ is the particle number per site (band filling), and
$a$ is the lattice spacing. We define the Fermi wave vector
$k_{f\sigma}\equiv\pi n_\sigma/a$ for each spin species $\sigma=\ua,\da$, and the FFLO
wave vector $q_\star\equiv k_{f\ua}-k_{f\da}$. The Hubbard model can be viewed as
the (regularized) lattice version of the continuum Gaudin-Yang model, with $a$ playing the role of
the short distance cutoff. Explicitly, the Gaudin-Yang model corresponds to
$H_U$ in the continuum limit, $a\rightarrow 0$,
for fixed particle numbers and system length $L$, i.e., $n\ll 1$.
In this limit, the parameters of $H_{GY}$ are related to those of $H_U$ by \cite{essler_book}
\be
m=\frac{\hbar^2}{2ta^2}, \;\;\; g=-Ua, \;\;\; \gamma=\frac{U}{2tn}. \label{cont}
\ee

Finally, the experimental
setup \cite{hulet} searching for quasi-1D FFLO state consists of a 2D
square lattice array of 1D tubes, { each described by} $H_{GY}$ and
coupled to its nearest neighbors by transverse tunneling of amplitude $t_\perp$,
\be
H_\perp=-t_\perp \int dx \sum_{\langle i,j
  \rangle,\nu,\sigma}[\psi^\dagger_{i,\nu,\sigma}(x)
  \psi_{j,\nu,\sigma}(x)+h.c.].
\ee
Here $\psi_{\nu=R/L,\sigma}$ is the fermion field operator for right/left
movers with spin $\sigma=\uparrow,\downarrow$ \cite{giamarchi}, and $\langle i,j\rangle$ labels nearest neighbor
tubes. Also relevant to our discussion is the intertube Josephson coupling of the form
\be
H_J=-J \int dx \sum_{\langle
  i,j\rangle}[\Delta^\dagger_{i}(x)\Delta_{j}(x)+h.c.].
\ee
The pair operator is defined as $\Delta(x)=\psi_{R,\ua}(x)\psi_{L,\da}(x)$, and
the pair tunneling, $J\propto t^2_\perp$, is generated by second order hopping process.
Intertube coupling establishes transverse coherence between
tubes and drives the system through a 1D-to-3D dimensional
crossover \cite{schulz}.

\section{Spin-charge mixing in an imbalanced Fermi gas}

We first obtain a low energy effective theory for the Gaudin-Yang model using standard Abelian
bosonization \cite{giamarchi,bible}, which is valid for weak interaction (small $\gamma$).
We linearize the spectrum of each spin species
around its Fermi points $\pm k_{f\sigma}$. Note that the two Fermi momenta differ by $\delta
k_f=k_{f\ua}-k_{f\da}$, since $N_\ua\neq N_\da$. We also define the
average and difference of Fermi velocities,
$\bar{v}_f=(v_{f\ua}+v_{f\da})/2$ and $\delta
v_f=v_{f\ua}-v_{f\da}$. We follow the notation convention of
Ref. \cite{giamarchi}, where the bosonization identity takes the form
\begin{align}
\psi_{R,\sigma}(x)=\frac{U_{R,\sigma}}{\sqrt{2\pi\alpha}}e^{ik_{f,\sigma}x}e^{-i[\phi_\sigma-\theta_\sigma]},\nn \\
\psi_{L,\sigma}(x)=\frac{U_{L,\sigma}}{\sqrt{2\pi\alpha}}e^{-ik_{f,\sigma}x}e^{i[\phi_\sigma+\theta_\sigma]}. \label{iden}
\end{align}
Here $\theta_{\sigma}$ is the dual field of boson field $\phi_{\sigma}$, $U_{R/L,\sigma}$ are the Klein factors, and
$\alpha$ is the short distance cutoff. The bosonized Gaudin-Yang model
\eqref{gaudin-yang} becomes,
\begin{align}
H_B&=\int \frac{dx}{2\pi}\left [ \bar{v}_f (\nabla\theta_c)^2 +
  \bar{v}^{+}_{f}(\nabla\phi_c)^2 +\delta v_f\delta k_f
  \frac{\nabla\phi_c}{\sqrt{2}}  \right. \nn \\
&+ \bar{v}_f (\nabla\tilde{\theta}_s)^2 +
  \bar{v}^-_f(\nabla\tilde{\phi}_s)^2 + \sqrt{2} \bar{v}^-_f\delta k_f
  \nabla\tilde{\phi}_s\nn \\
&+ \delta v_f (\nabla\theta_c\nabla\tilde{\theta}_s+\nabla
  \phi_c\nabla \tilde{\phi}_s)
+ \left. \frac{g}{\pi\alpha^2}\cos(\sqrt{8}\tilde{\phi}_s) \right
  ]. \label{bosonized_GY}
\end{align}
Here the charge and spin
field $\phi_{c / s}=(\phi_\ua\pm \phi_\da)/\sqrt{2}$ are defined in
the standard way. We have introduced
 shifted spin fields \[\tilde{\phi}_s = \phi_s-\delta k_f
x/\sqrt{2},\;\; \tilde{\theta}_s=\theta_s,\] and a short hand notation
$\bar{v}^\pm_f=\bar{v}_f\pm g/\pi$.

For the unpolarized case, $p=0$, both $\delta k$ and $\delta v_f$ vanish, so
$H_B$ is reduced to the sine-Gordon model. With a { negative
  coefficient ($g<0$),}
the cosine term is relevant. It is well known
that the system, an attractive Fermi gas, is a Luther-Emery liquid where spin and charge separation holds:
the charge sector is gapless and described by a Gaussian Hamiltonian and the
spin sector is gapped. The algebraic decay of the $s$-wave paring susceptibility, which is most diverging,
is determined by the Luttinger parameter in the charge sector.

Finite population imbalance brings several significant differences.
First, it introduces a linear ``source" term for the (shifted) spin field which acts as
an {\it effective} magnetic field,
\[
h=\sqrt{2}
\bar{v}^-_f \delta k_f.
\]
But more importantly, it leads to coupling between the spin and charge
sector, described by the density and current interaction terms ($\nabla\theta_c\nabla\tilde{\theta}_s$ and $\nabla
  \phi_c\nabla \tilde{\phi}_s$) of order $\delta v_f$ in $H_B$.
Therefore, Eq.~\eqref{bosonized_GY} clearly shows that a 1D
attractive Fermi gas with finite $p$ is in general not a
spin-charge separated Luther-Emery liquid.

In the limit of very small $p$, both $\delta v_f$ and $\delta k_f$ are small so
we can neglect high order terms $\propto \delta v_f \delta k_f$ in
Eq.~\eqref{bosonized_GY}. Furthermore, in situations where we are allowed to neglect the spin-charge mixing terms
$\propto \delta v_f$ but keep the linear $\nabla\tilde{\phi}_s$ term
$\propto\delta k_f$, then Eq.~\eqref{bosonized_GY} reduces to the
spin-charge separated model proposed by Yang to describe quasi-1D superconductors in magnetic field \cite{kyang},
\begin{align}
H_{Y}&=\int \frac{dx}{2\pi}\left [ \bar{v}_f (\nabla\theta_c)^2 +
  \bar{v}^{+}_{f}(\nabla\phi_c)^2  \right. \nn \\
&+ \bar{v}_f (\nabla\tilde{\theta}_s)^2 +
  \bar{v}^-_f(\nabla\tilde{\phi}_s)^2 + h
  \nabla\tilde{\phi}_s\nn + \left. \frac{g}{\pi\alpha^2}\cos(\sqrt{8}\tilde{\phi}_s) \right
  ].
\end{align}
Its charge sector is a massless Gaussian, while its spin sector (the second line) is a sine-Gordon
Hamiltonian in an {\it effective} magnetic field $h$, which we call $H_{sGh}$. $H_{sGh}$ has
been investigated thoroughly by many authors since the pioneer works of Ref. \onlinecite{haldane,japa}.
Increasing $h$ over a critical value $h_c$ will trigger a phase transition from a Luther-Emery
liquid (BCS-like superfluid with algebraic order and spin gap) to a FFLO-like phase in which the
spin excitation is also massless and the pair correlation function oscillates at wave vector $q_\star$
on top of the algebraic decay. The spin sector of the 1D FFLO phase of $H_Y$ can be described
as a Luttinger liquid of spin solitons \cite{haldane}. These solitons are the 1D counterpart
of the static order parameter domain walls (more precisely, the Andreev bound states at the domain walls populated by the majority spins \cite{parish}) which form a crystal in high dimensional FFLO phases \cite{rainer}.

Based the spin-charge separated Hamiltonian $H_Y$, Yang also pointed out that the magnetic
field driven BCS-FFLO transition is in the universality class of
  2-dimensional classical commensurate-incommensurate
transitions \cite{kyang}. To see this clearly, we can solve $H_{sGh}$ analytically along the Luther-Emery line to map it onto a free (massive) fermion Hamiltonian by
refermionization \cite{giamarchi,bible}:
the magnetic field $h$ then plays the role of chemical potential, while $g$ controls the size of the
band gap. As $h$ is increased to $h_c$, the Zeeman energy exceeds the band gap,
so the upper band becomes populated by spinless fermions (spin solitons).
This general picture holds even away from the Luther-Emery line, where spin solitons become
interacting with each other. Thus, Yang's model $H_{Y}$ gives
an appealing physical picture for the 1D FFLO like phase in terms
of solition liquids, provided that the basic assumption of spin-charge separation holds.

Yet, the spin-charge mixing terms in $H_B$ have scaling dimensions of $2$ and
are marginal operators in the renormalization group sense. So they cannot
be simply neglected even if their values are small. Moreover, in
cold atom experiments the population imbalance $0\leq p\leq 1$ can be arbitrary
and usually not small. This is drastically
different from typical situations in solids where the Zeeman energy is
much smaller than the Fermi energy. In general, the spin-charge mixing
terms have to be treated on the same footing as other terms in $H_B$.

While a direct perturbative analysis of $H_{B}$ is possible, it is unclear
how well the bosonized Hamiltonian $H_B$ describes the system for the case of
strong interaction and large population imbalance.
For these reasons, we shall adopt an alternative but much more powerful
method, which is valid for arbitrary $(n, p, g)$ and treats the spin-charge mixing effect exactly,
to construct the bosonized effective
field theory of $H_{GY}$. The method is based on the Bethe ansatz solution
of the microscopic model $H_{GY}$ (or $H_{U}$) and general principles
of conformal field theory.

\section{Effective theory for the 1D FFLO phase}
\label{sec:effective_theory}

In this section, we show that despite the spin-charge mixing, the 1D
FFLO phase features a more general decoupling of two critical degrees
of freedom.  In the language of conformal field theory, the 1D FFLO
state is a critical phase with two gapless normal modes and it is
formally described by the direct product of two Virasoro algebras with
central charge $c=1$ \cite{frahm}. Such a ``two fluid" description is
in accordance to our physical intuition in higher dimensions, e.g.,
the 3D attractive Fermi gas with population imbalance in the strong
interaction (BEC) limit can be described by a fermion-boson mixture
with some residue interactions. However, the distinction between
bosons and fermions is lost in 1D.  Our task is to identify the two
normal modes (elementary excitations) and establish their relation
with the usual spin/charge excitations.

\subsection{Summary of main results}
We first present the final results, so readers not interested in
technical details may skip the derivation and proceed directly to
the next section. The 1D FFLO state is a two component Luttinger
liquid described in bosonized form by the effective Hamiltonian
\be
H_{\mathrm{FFLO}} = \sum_{i=1,2} \int \frac{dx}{2\pi} u_i \left[
  (\nabla \vartheta_i)^2 + (\nabla \varphi_i)^2
  \right]. \label{luttinger2}
\ee
The two normal modes ($i=1,2$) are decoupled and have
different group velocities, $u_1$ and $u_2$. Each mode, described by
the boson field ${\varphi_i}$ and its dual $\vartheta_i$, is a
superposition of the spin up and down fields $\phi_\sigma$ and
$\theta_\sigma$,
\be
\left(\begin{array}{c}
                \varphi_1\\
               \varphi_2
\end{array}\right)
=(\bar{Z}^{\mathrm{T}})^{-1}
\left(\begin{array}{r}
               -\phi_\ua\\
               \phi_\da
\end{array}\right)
,\;
\left(\begin{array}{c}
                \vartheta_1\\
               \vartheta_2
\end{array}\right)
=\bar{Z}
\left(\begin{array}{r}
                -\theta_\ua\\
              \theta_\da
\end{array}\right).
\label{transform}
\ee
The superscript T means matrix transpose, and the so-called
dressed charge matrix is given by
\be
\bar{Z}=\left[
\begin{array}{lr}
               Z_{cc}-Z_{sc} & Z_{sc}     \\
               Z_{ss}-Z_{cs} & -Z_{ss}
\end{array}
\right]. \label{Zbar}
\ee
Eq.~\eqref{transform} clearly shows that in general each normal mode
$\varphi_i$ is a mixture of the spin field $\phi_s$ and the charge
field $\phi_c$. Therefore, elementary excitations of the FFLO state
are fractional states carrying both charge and spin, i.e., the
superposition of spinons and holons \cite{pham}.  The degree of
spin-charge mixing is determined by the dressed charge matrix
$\bar{Z}$, or equivalently matrix
$\bar{Y}\equiv[\bar{Z}^{\mathrm{T}}]^{-1}$
\cite{pham,frahm,vekua,furusaki}.  As shown in Ref. \cite{pham},
spin-charge separation occurs only if the charge matrix
\[\hat{g}\equiv\bar{Z}^{-1}\bar{Y}\] possesses a $Z_2$ symmetry,
$g_{11}=g_{22}$, $g_{12}=g_{21}$, i.e., $\hat{g}$ is invariant under
the exchange $1\leftrightarrow 2$. All the parameters in the effective
bosonized Hamiltonian $H_{\mathrm{FFLO}}$, such as
$\{\bar{Z}_{ij},u_i\}$, can be computed from the parameters of the
microscopic Hamiltonian $H_{GY}$ or $H_U$.
Eq.~\eqref{luttinger2}-\eqref{Zbar} are the main results of this
section, they form the basis for all subsequent discussions on
correlation functions and quasi-1D phases.

\subsection{The method}
\label{sec:steps}
Below we outline the main steps leading to the effective
field theory Eq.~\eqref{luttinger2} and the algorithm to compute
parameters $\{\bar{Z}_{ij},u_i\}$. The procedure takes a detour
to make use of several well established, nontrivial results for the
1D Hubbard model.

(I) We start from the attractive Hubbard model $H_U$. Its continuum limit,
summarized in Eq.~\eqref{cont}, yields the Gaudin-Yang model $H_{GY}$. Therefore,
in this limit, the { effective field
theory of} the two models are the same. Indeed, there is a one to
one correspondence between the Bethe ansatz integral equation for the
Hubbard model \cite{essler_book} and the Gaudin integral equation for
Gaudin-Yang model \cite{taka}. In principal, all analysis can be performed on
the Gaudin-Yang model directly without resorting to its lattice
regularization to arrive at the same conclusion. In practice, however,
the latter route is less economical since the Bethe ansatz solution of
the continuum model, especially its dressed charge formalism, is less
developed.

(II) We map the attractive Hubbard model $H_U$ for given $(U/t,n,p)$
to a repulsive Hubbard model $H'_{U}$ with onsite repulsion $U$,
density $n'=1-np$, and imbalance $p'=(1-n)/(1-np)$ by the well-known
staggered particle-hole transformation \cite{essler_book,woy2,emery,luscher},
\be
c_{i,\ua}\rightarrow (-1)^ic^\dagger_{i,\ua},\;\;
c_{i,\da}\rightarrow c_{i,\da}.
\ee
The motivation
behind this is two-fold. First, it is technically more convenient to
solve $H'_U$ via Bethe ansatz \cite{woy2}. Second, as shown by Penc and
S\'olyom \cite{penc} based on the work of Frahm and Korepin \cite{frahm}, the
repulsive Hubbard model at general (non-half) filling and magnetic field is
equivalent to a two-component Tomonaga-Luttinger (TL) liquid.
We shall exploit this result to construct the field theory
for $H_U$ (and $H_{GY}$) from the Bethe ansatz solution of
$H'_U$.

(III) For given $(U/t,n',p')$, we numerically solve the coupled Bethe
ansatz integral equations for $H'_U$. We first determine the density
distribution functions $\rho(k)$ and $\sigma(\lambda)$, defined within
the Fermi interval $k\in(-Q,Q)$ and spin rapidity range
$\lambda\in(-A,A)$ respectively. This requires solving
\begin{align*}
\rho (k)&=\frac{1}{2\pi}+\cos k \int_{-A}^{A} d \lambda a_1(\sin k-\lambda)\sigma(\lambda),\\
\sigma(\lambda)&=\int_{-Q}^{Q} dk \rho (k) a_1(\sin k-\lambda) \\
&-\int_{-A}^{A} d \lambda' \sigma(\lambda') a_2(\lambda-\lambda'),
\end{align*}
where $a_n(x)=nU/(4\pi t)/[(U/4t)^2+x^2]$, together with the number constraint equations,
\begin{align*}
\int_{-Q}^Q dk \rho (k)&=n',\\
\int_{-A}^{A} d\lambda \sigma(\lambda)&=n'_\da=\frac{n'}{2}(1-p').
\end{align*}
This is done by an iteration procedure which also fixes $Q$ and $A$
self-consistently. Each integral equation is converted to an
algebraic equation using Gaussian-Legendre quadrature. The code was
checked to yield exactly the same result as
those in Ref. \cite{frahm,penc}.

Once $Q$ and $A$ are found, it is then straightforward to find other quantities: the
dressed energies $[\epsilon_c(k),\epsilon_s(\lambda)]$, their derivatives,
$[\epsilon'_c(k),\epsilon'_s(\lambda)]$, and the dressed charges,
$[\xi_{cc}(k), \xi_{cs}(\lambda), \xi_{sc}(k), \xi_{ss}(\lambda)$], by
solving corresponding integral equations. These lengthy
equations are well documented and will not be repeated here \cite{frahm,penc,essler_book}.
The dressed charge matrix elements are given by
\[
\begin{array}{cc}
            Z_{cc}=\xi_{cc}(Q), & Z_{cs}=\xi_{cs}(A), \\
            Z_{sc}=\xi_{sc}(Q), & Z_{ss}=\xi_{ss}(A).
\end{array}
\]
The normal mode group velocities are,
\[
u_1=\frac{\epsilon'_c(Q)}{2\pi \rho(Q)},\;u_2=\frac{\epsilon'_s(A)}{2\pi \sigma(A)}.
\]
Note that the subscripts $c$ and $s$ here and in the Bethe ansatz literature have to be distinguished
from the charge and spin index in Abelian bosonization such as in Eq.~\eqref{bosonized_GY}.
This fact has been emphasized for example by Voit et al \cite{voit}. To avoid confusion, we adopt index $i=1,2$
to label the two normal modes instead of $c$ and $s$.

(IV) We map $H'_U$ to a two-component Tomonaga-Luttinger Hamiltonian, $H'_{TL}$, following
 Penc and S\'olyom \cite{penc}. It becomes transparent in bosonized form
\begin{align*}
H'_{TL}=\int \frac{dx}{2\pi}\left [ (v_{f,\ua}+\tilde{g}_{4,\ua}-\tilde{g}_{2,\ua}) (\nabla\theta_\ua)^2  \right. \\
+ (v_{f,\ua}+\tilde{g}_{4,\ua}+\tilde{g}_{2,\ua}) (\nabla\phi_\ua)^2  \\
+ (v_{f,\da}+\tilde{g}_{4,\da}-\tilde{g}_{2,\da}) (\nabla\theta_\da)^2  \\
+ (v_{f,\da}+\tilde{g}_{4,\da}+\tilde{g}_{2,\da}) (\nabla\phi_\da)^2  \\
+  \left. 2(\tilde{g}_{4,\perp}- \tilde{g}_{2,\perp}) \nabla\theta_\ua\nabla \theta_\da
+  2(\tilde{g}_{4,\perp}+ \tilde{g}_{2,\perp}) \nabla\phi_\ua\nabla\phi_\da \right].
\end{align*}
The Luttinger parameters in the g-ology notation, $\tilde{g}$, are given in terms of $\{u_i,Z_{ij}\}$
by Eq.~(6.12) to (6.17) in Ref. \onlinecite{penc}. For example,
\begin{align*}
v_{f,\ua}+\tilde{g}_{4,\ua}-\tilde{g}_{2,\ua}&=u_1(z_{cc}-Z_{sc})^2 + u_2 (Z_{cs}-Z_{ss})^2, \\
v_{f,\ua}+\tilde{g}_{4,\ua}+\tilde{g}_{2,\ua}&=\frac{u_1Z^2_{ss}+ u_2 Z^2_{sc}}{(Z_{cc}Z_{ss}-Z_{cs}Z_{sc})^2}.
\end{align*}
The low energy effective theory $H'_{TL}$ yields the same finite
size spectrum and correlation functions as the original microscopic
model $H'_U$. It is instructive to compare this nonperturbative
result with that obtained by weak coupling bosonization, $H_B$ in Eq.,
\eqref{bosonized_GY}.

(V) Finally we particle-hole transform $H'_{TL}$ back to obtain the
effective Hamiltonian $H_{TL}$ for $H_U$. In bosonized form, $H_{TL}$ is
obtained from $H'_{TL}$ by replacement $\phi_\ua\rightarrow -\phi_\ua$
and $\theta_\ua\rightarrow - \theta_\ua$. Since $H_{TL}$ is quadratic,
we diagonalize it by a canonical transformation
using matrix $\bar{Z}$ \cite{pham,vekua}.  The end result is
Eq.~\eqref{luttinger2}.

\section{Single particle and pair correlation functions}
\label{sec:single+pair}

The bosonized effective Hamiltonian Eq.~\eqref{luttinger2} makes it
straightforward to compute the correlation function of any operator of
 the form
\[
O(x)=e^{i[\hat{a}^\mathrm{T} \hat{\phi}(x) + \hat{b}^\mathrm{T} \hat{\theta}(x)]}
\]
where $\hat{a}$ and $\hat{b}$ are arbitrary two-component vectors, and
$\hat{\phi}^\mathrm{T}=(\phi_\ua,\phi_\da)$,
$\hat{\theta}^\mathrm{T}=(\theta_\ua,\theta_\da)$.  Using the
relation Eq.~\eqref{transform}, we find that the correlation function  $\langle O(x)O^\dagger(0)\rangle$ has the
asymptotic behavior of $x^{-2\delta}$ at zero temperature as $x\rightarrow \infty$. The scaling dimension of $O$ is given by
\[
\delta = \frac{1}{4}[\hat{a}^\mathrm{T}\hat{g}^{-1}\hat{a}
+\hat{b}^\mathrm{T}\hat{g}\hat{b}].
\]
The charge matrix is defined by $\hat{g}=\bar{Z}^{-1}\bar{Y}$.

We apply this result to compute the single particle propagator and the singlet pair susceptibility
\begin{align}
G_\sigma(x,\tau)&=-\langle
T_\tau\psi_{R,\sigma}(x,\tau)\psi_{R,\sigma}^\dagger(0,0)\rangle, \nn
\\
\chi_(x,\tau)&=-\langle T_\tau \Delta(x,\tau)\Delta^\dagger(0,0)\rangle .
\end{align}
We shall focus on $G_\ua$ for the majority spin ($\ua$),
since $G_\da$ only has subdominant power law divergence (its scaling dimension
is larger than that of $G_\ua$, i.e., $\delta_\da > \delta_\ua$).
For example, from the bosonized representation of fermion field operator $\psi_{R,\sigma}$,
Eq.~\eqref{iden}, we find $a=(-1,0)^\mathrm{T}$ and $b=(1,0)^\mathrm{T}$ for $G_\ua$.
It follows that for $x\rightarrow \infty$ at zero
temperature,
\begin{align}
G_\ua(x)&\sim \frac{e^{ik_{f\ua} x} }{x^{2\delta_\ua}},\;\; \delta_{\ua}=\beta_1+\beta_2,\\
\chi (x)&\sim
\frac{e^{iq_\star x}}{x^{2\delta_\Delta}}, \;\; \delta_\Delta=\gamma_1+\gamma_2.\label{asymp}
\end{align}
The scaling dimensions of the single particle field operator $\psi_\ua$ and pair operator $\Delta$
are directly related to the dress charge matrix elements that describe the effect of spin-charge mixing,
\begin{align}
\beta_1&=\frac{1}{4}[(\bar{Z}_{11})^2+(\bar{Y}_{11})^2], &\beta_2=\frac{1}{4}[(\bar{Z}_{21})^2+(\bar{Y}_{21})^2]; \nn\\
\gamma_1&=\frac{1}{4}[Z^2_{cc}+\frac{Z^2_{cs}}{(\det\bar{Z})^2}],
&\gamma_2=\frac{1}{4}[Z^2_{cs}+\frac{Z^2_{cc}}{(\det\bar{Z})^2}].
\label{col}
\end{align}
Here $\bar{Z}_{11}$ denotes the (1,1) element of the matrix
$\bar{Z}$. We also define the corresponding anomalous dimensions
$\eta_\ua=2-2\delta_\ua$, and $\eta_\Delta=2-2\delta_\Delta$.

Eq.~\eqref{asymp} is the hallmark of 1D FFLO phase: the $s$-wave
pair correlation function oscillates in space at the FFLO wave vector $q_\star=k_{f,\ua}-k_{f,\da}$
on top of the algebraic decay. Although it looks similar to
Yang's conclusion based the spin-charge separated model $H_Y$, the present
result fully takes into account the effect of spin-charge mixing
and remains valid for all values of $0<p<1$. Our results can also shed
more light on the nature of the magnetic field driven BCS-FFLO transition
in 1D. Under a particle-hole transformation, it is equivalent to the Mott
transition of repulsive Fermi gas in magnetic field (driven by varying chemical potential).
As shown recently by Frahm and Vekua \cite{vekua}, due to
the intrinsic spin-charge coupling, it
is {\it not} in the same universality class  as the single mode commensurate-incommensuarate
transition.

\begin{figure}[htb]
\includegraphics[width=2.5in]{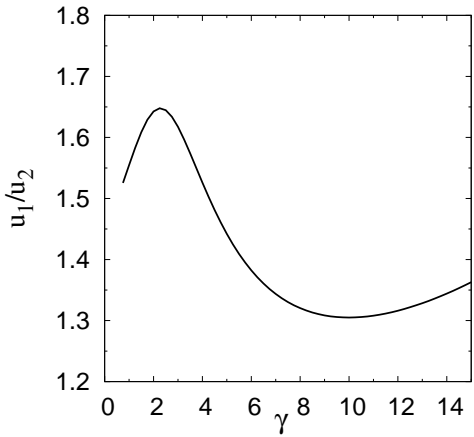}
\caption{The ratio $u_1/u_2$ as function of $\gamma$ for $p=0.2$. }
\label{uratio}
\end{figure}

At finite temperature $T$, $G_\ua(x,\tau)$ and $\chi(x,\tau)$ still
factorize due to the
decoupling of two normal modes. After some algebra, we obtain the Fourier
component $G_0=G_{\ua}(k=k_{f,\ua},\omega=0)$ (up to a phase factor),
\be
G_0= ( 2\pi^3 \alpha T^2)^{-1}\, \mathrm{I}_1(\frac{u_1}{u_2}) \prod_{i=1,2} (\frac{\pi T\alpha}{u_i})^{2\beta_i} \sqrt{u_i}. \label{g0}
\ee
Similarly, $\chi_0=\chi(k=q_\star,\omega=0)$ is found to be
\be
\chi_0 = ( 2\pi^2\alpha T)^{-2}\, \mathrm{I}_2(\frac{u_1}{u_2}) \prod_{i=1,2} (\frac{\pi T\alpha}{u_i})^{2\gamma_i} \sqrt{u_i}. \label{chi0}
\ee
Here, the dimensionless function I$_{1/2}$ are defined as
\begin{align*}
\mathrm{I}_1(y)&=\int_{-\infty}^{\infty} dx\int_0^\pi dt\sin t\prod_{i} [\sinh^2(x y^{i-\frac{3}{2}})+\sin^2(t) ]^{-\beta_i}, \\
\mathrm{I}_2(y)&=\int_{-\infty}^{\infty} dx\int_0^\pi dt \prod_{i} [\sinh^2(x y^{i-\frac{3}{2}})+\sin^2(t) ]^{-\gamma_i}.
\end{align*}
These results will become useful in the RPA calculation for the quasi-1D systems.
Note that the ratio $u_1/u_2$ is not a rapidly changing function of $\gamma$. Fig. \ref{uratio}
shows its value for $p=0.2$.

\section{Phase diagram of the quasi-1D system at $T=0$.}

We now apply the effective theory for single tube to study the
quasi-1D system realized in experiments: a square lattice array of
such tubes coupled by transverse hopping $t_\perp$. We are interested
in the fate of the 1D FFLO phase as the $t_\perp$ is turned on.

In the limit of small $t_\perp$, we can treat $H_\perp$ and
$H_J$ as perturbations to the individual 1D Hamiltonian $H_{\mathrm{FFLO}}$. In
the language of renormalization group (RG), both $H_\perp$ and $H_J$
are relevant perturbations: the scaling dimensions of the single
particle and pair tunneling
are both smaller than $2$. The first-order RG equations
for the effective inter-tube couplings
$t_\perp(\kappa)$ and $J(\kappa)$ at momentum scale $\kappa$ (with
$\kappa\rightarrow 0$) read:
\be
\begin{array}{c}
\kappa dt_\perp(\kappa)/d\kappa =(2\delta_\ua-2)t_\perp(\kappa),
\\
\kappa dJ(\kappa)/d\kappa =(2\delta_\Delta-2) J(\kappa),
\end{array}
\ee
where the equation of $t_\perp$ is for the spin
$\ua$ tunneling (as we have emphasized before, it is more relevant than the spin $\da$). In other words,
the fate of the 1D FFLO phase is controlled by the
relative magnitude $\delta_\ua$ and
$\delta_\Delta$ \cite{kyang}. For
$\delta_\Delta<\delta_\ua$, pair tunneling is most relevant and the
system flows into a quasi-1D
FFLO state. In this state, strong effective Josephson coupling locks the phases of all tubes to
establish the overall phase coherence to produce a genuine superfluid state. For $\delta_\Delta>\delta_\ua$,
however, single particle tunneling is most relevant and the system flows into a
partially polarized Fermi liquid (FL) state with well defined quasiparticles.
In the latter case,
the actual ground state of the quasi-1D system at zero temperature
depends on the residue interactions between quasiparticles, the
details of which are not captured by the leading order RG analysis presented
here \cite{kyang}. Such limited predictive power is inherent to all leading
order RG analysis on coupled Luttinger liquids.
Therefore, the Fermi liquid state predicted here should be understood
as a region in the phase diagram where the superfluid transition temperature
is significantly suppressed by the weakening of effective
intertube Josephson coupling. It is important to bear in mind that other
instabilities may take over at lower temperatures leading to a ground
state with broken symmetry.
\begin{figure}[htb]
\includegraphics[width=2.5in]{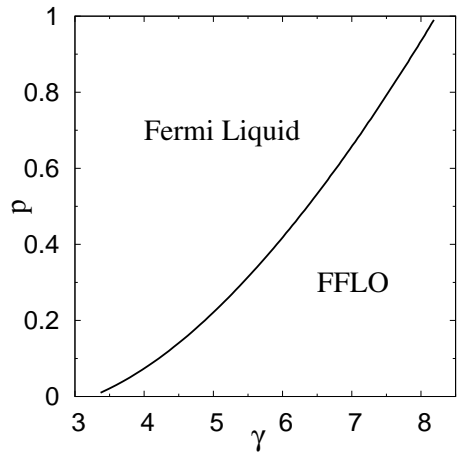}
\caption{Zero temperature phase diagram of a
quasi-1D attractive Fermi gas in the limit of weak inter-tube
tunneling. $\gamma$ is the interaction
strength of the Gaudin-Yang model and $p$ is the population imbalance.}
\label{f1}
\end{figure}

\begin{figure}[htbp]
\includegraphics[width=2.5in]{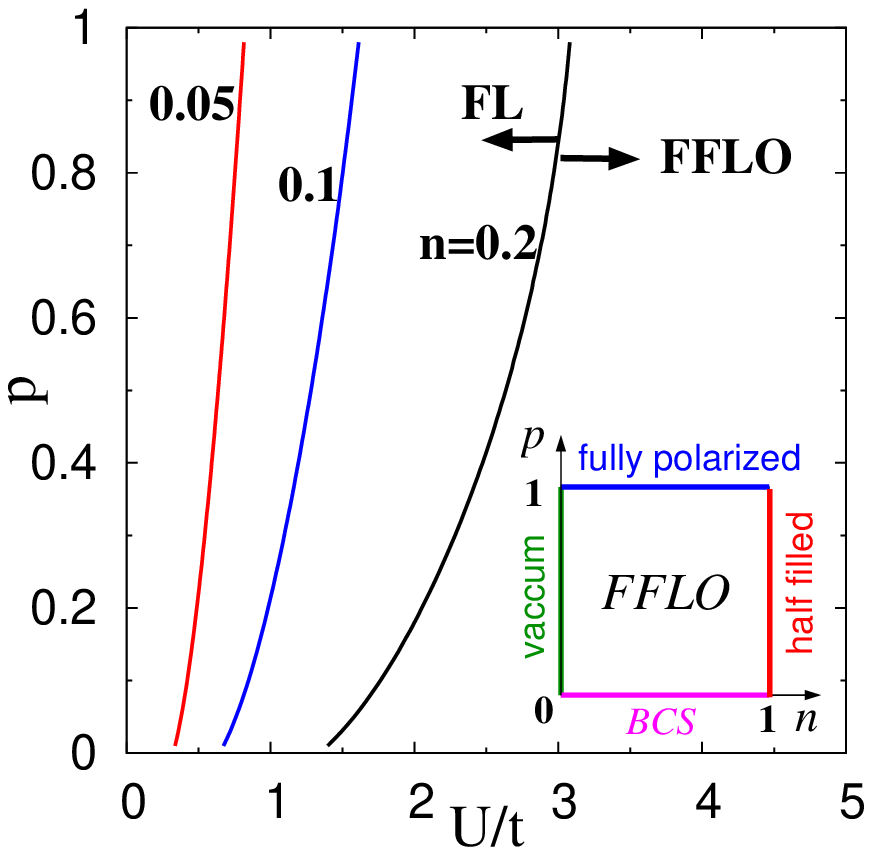}

\includegraphics[width=2.5in]{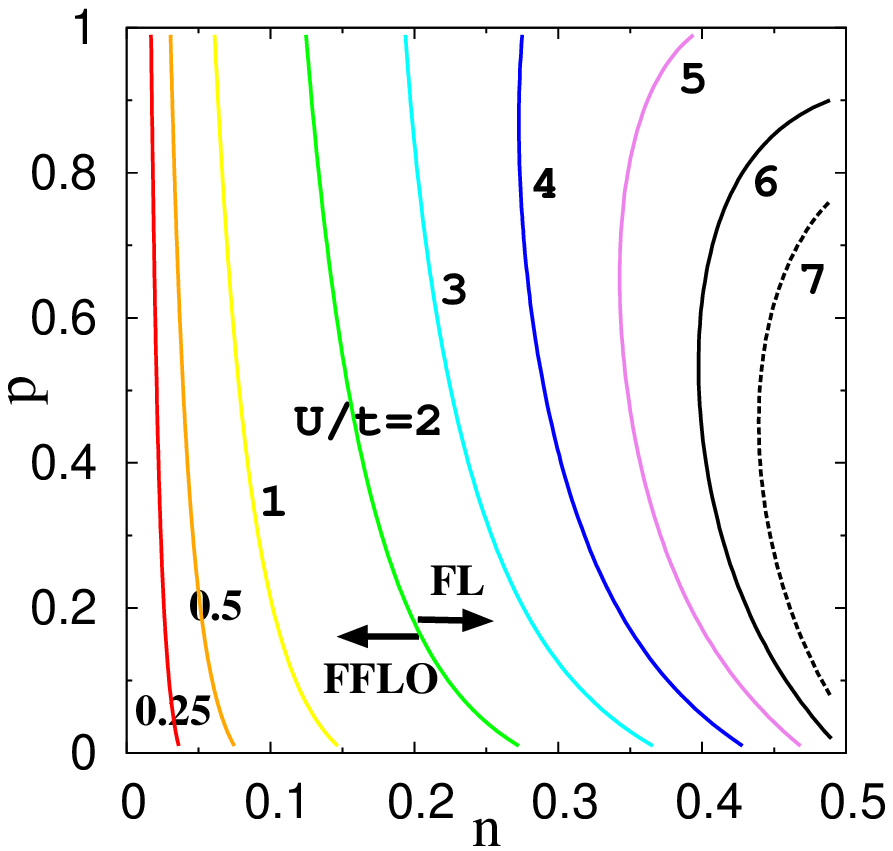}
\caption{(Color online) Zero temperature phase diagram of the quasi-1D
  attractive Hubbard model, $H_U+H_\perp+H_J$, in the limit of small
  $t_\perp$. $p$ is the population imbalance. Each line for fixed
  density $n$ (top panel) or interaction strength $U/t$ (bottom panel)
  is the phase boundary separating the FFLO and
  Fermi liquid (FL) phase (indicated by arrows). Inset: phase diagram
  of the 1D attractive Hubbard model (single tube).  }
  \label{phase-diagram}
\end{figure}

The $T=0$ phase diagram of the quasi-1D gas system based on leading
order RG is shown in Fig. \ref{f1}.
It is obtained by going
through the steps outlined in Sec.\ref{sec:steps} and taking
the continuum limit, $a\rightarrow 0$,
for fixed particle numbers and system length $L$ [see Eq.~\eqref{cont}].
As the intertube
tunneling is turned on, the 1D FFLO phase (originally occupying the
whole region $0<p<1$ for all $\gamma$) splits into two distinct
phases, an FFLO superfluid and a polarized Fermi liquid (FL).
Intuitively, stronger attractive interaction (larger $\gamma$) favors
the FFLO phase. From Fig. \ref{f1}, one can read off the critical
interaction strength required to realize the quasi-1D FFLO state for
given $p$. For fixed interaction $\gamma$, increasing imbalance would
drive the system out of FFLO into a Fermi liquid phase.  A crucial
feature of the phase diagram is that the FFLO phase survives in a
smaller region in quasi-1D than true 1D (single tube).  This shrinking
trend observed at $t_\perp\rightarrow 0$ is expected to continue as
$t_\perp$ increases, since we know that FFLO state in 3D only occupies
a tiny part of the phase diagram \cite{sheehy}.
Fig. \ref{phase-diagram} shows $T=0$ phase diagram of the quasi-1D
attractive Hubbard model in the limit of $t_\perp\rightarrow 0$.  We
observe a similar splitting of the 1D FFLO phase. 
Ref. \onlinecite{luscher} also discussed the phases of weakly coupled Hubbard
chains in the presence of finite spin polarization, close in spirit with ours. 
While our main focus here is the continuous gas systems (without lattice in the 
$x$ direction), our result of quasi-1D Hubbard model (Figure 3) agrees with 
Ref. \onlinecite{luscher}.

\section{The transition temperature of quasi-1D FFLO state}

In this section, we use the random phase approximation (RPA)
\cite{schulz,wen,boies} to compute the FFLO superfluid transition
temperature $T_c$ for the quasi-1D system with a small but finite
$t_\perp$, going beyond the perturbative RG analysis in the previous
section. RPA has been successfully applied to study coupled Luttinger
liquids ($p=0$) \cite{boies}. Here we generalize it to the case of
$p>0$ with spin-charge mixing. Within RPA, the 3D single particle
propagator
\be
\mathscr{G}^{-1}_{\ua}(k,k_\perp=0,\omega_n)=G^{-1}_{\ua}(k,\omega_n)-z_\perp
t_\perp, \label{rpa1}
\ee
and the 3D pair susceptibility
\be
\mathscr{X}^{-1}(q_\star,k_\perp=0,\omega=0)=\chi_0^{-1} - z_\perp J,
\ee
where $k_\perp$ is the transverse momentum and $z_\perp=4$ is the
{ transverse coordination number of every single atomic gas tube in a
square lattice array.} The 1D propagator $G_{\da}$ and
susceptibility $\chi_0$ have been computed in Section
\ref{sec:single+pair} and are given by Eq.~\eqref{g0} and \eqref{chi0},
respectively.

\begin{figure}[htbp]
\includegraphics[width=2.5in]{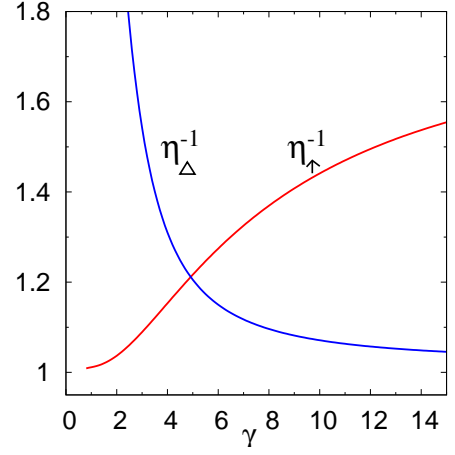}
\caption{RPA predictions for the
$t_\perp$ dependence of the single particle crossover (Luttinger
liquid to Fermi liquid) temperature, $T_{\mathrm{FL}}\sim
t_\perp^{1/\eta_\ua}$, and the FFLO transition temperature, $T_c\sim
t_\perp^{2/\eta_\Delta}$, as functions of $\gamma$ for $p=0.2$. }
\label{rpa}
\end{figure}

RPA predicts two temperature scales that characterize the 1D-to-3D
crossover as the temperature is lowered.
$\mathscr{G}_\ua(k_{f\ua},0,\omega_{n=1})$
starts to develop a pole at the single particle crossover temperature
$T_{\rm FL}$. From Eq.~\eqref{rpa1}, we find
\be
T_{\rm FL}\sim W\left[\frac{t_\perp z_\perp}{W}
  \mathrm{I}_1(\frac{u_1}{u_2})\right]^{1/\eta_\ua}
\prod_i\left(\frac{u_i}{\bar{v}_f}\right)^{({1/2-2\beta_i})/{\eta_\ua
}}. \label{t1x}
\ee
Similarly, the divergence of the 3D pair susceptibility $\mathscr{X}(q_\star,0,0)$ defines the two
particle crossover temperature $T_c$,
\be
T_c \sim W \left[\frac{Jz_\perp
  }{\bar{v}_f}\mathrm{I}_2(\frac{u_1}{u_2})\right]^{1/\eta_\Delta}
\prod_i \left(\frac{u_i}{\bar{v}_f}\right)^{
  ({1/2-2\gamma_i})/{\eta_\Delta}} . \label{tc}
\ee
Here the ultraviolet cutoff $W=\bar{v}_f/\alpha$ is roughly the band
width (or Fermi energy). Above max$(T_{\rm FL},T_c)$, thermal fluctuation
destroys coherence between the tubes,
the system behaves as uncoupled two-component Luttinger liquids with
fractional excitations, each described by $H_{FFLO}$. As the temperature
is lowered, if $T_{\rm FL}>T_c$, these fractional excitations
first confine into sharply defined quasiparticles at temperature scale
$T_{\rm FL}$, and the system crosses over to a partially polarized Fermi liquid
(the Fermi liquid may develop other instabilities at lower temperature depending
on the details of the residue interactions between the quasiparticles).
On the other hand, if $T_c>T_{\rm FL}$,
the system first undergoes a phase transition into an FFLO phase
with long range order.
In this case, $T_c$ can be identified as the superfluid transition temperature.
In the limit of vanishing $p$, the transition into the FFLO superfluid
can be viewed as the condensation of soliton liquid into crystal, i.e. static
domain walls \cite{kyang}.

Whether the single or two particle crossover occurs first depends on
the interaction strength. At low densities, we find from
Eq.~\eqref{t1x} and \eqref{tc} that $T_{\rm FL}>T_c$ at small $\gamma$
(weak interaction) while $T_c>T_{\rm FL}$ for large $\gamma$ (strong
interaction). Thus, the schematic plot of $T_{\rm FL}$ and $T_c$ as
functions of $\gamma$ looks similar to Fig. 1 of Ref. \cite{boies} for
coupled Luttinger liquids. At some critical value of $\gamma$, $T_{\rm
FL}$ and $T_c$ are comparable to each other. This marks the phase
boundary between the Fermi liquid and the FFLO phase. All these
results are in qualitative agreement with those obtained in the
previous section from RG analysis.

Eq.~\eqref{tc} shows that $T_{c}$ scales with
$J$ as a power law, with the exponent given by the inverse of anomalous
dimension $\eta_\Delta$,
\[
T_c\propto J^{1/\eta_\Delta}.
\]
The exponents
$\eta^{-1}_\Delta$ and $\eta^{-1}_\ua$ are plotted in
Fig. \ref{rpa} for $p=0.2$. We
observe that at weak interaction $T_{\rm FL}\propto t_\perp$, while at strong
interaction $T_c\propto J \propto t^2_\perp$. The growth of $T_c$ with
inter-tube coupling will eventually stop when $t_\perp$ becomes so
large that it can no longer be treated as a perturbation. Then the system
becomes more 3D-like, and $T_c$ starts to drop as $t_\perp$ is further
increased \cite{parish}. Our results support the argument of
Ref. \cite{parish} that the optimal value of $T_c$ is realized for
small but finite $t_\perp$.

\begin{acknowledgments}
We thank M. A. Cazalilla, A. Ho, D. Huse, A. Paramekanti, and
especially R. Hulet for illuminating discussions.
This work is supported under ARO
Award No. W911NF-07-1-0464 with funds from the DARPA OLE Program and
ARO Award No. W911NF-07-1-0293.
\end{acknowledgments}


\begin{thebibliography}{10}

\bibitem{rice1}G. B. Partridge, W. Li, R. I. Kamar, Y.-A. Liao, and R. G. Hulet, Science 311, 503 (2006).
\bibitem{mit1}M. W. Zwierlein, A. Schirotzek, C. H. Schunck, and W. Ketterle, Science 311, 492 (2006).
\bibitem{mit2}M. W. Zwierlein, C. H. Schunck, A. Schirotzek, and W. Ketterle, Nature 442, 54 (2006).
\bibitem{shin}Y. Shin, M. W. Zwierlein, C. H. Schunck, A. Schirotzek, and W. Ketterle, Phys. Rev. Lett. 97, 030401 (2006).
\bibitem{rice2}G. B. Partridge, W. Li, Y. A. Liao, R. G. Hulet, M. Haque, and H. T. C. Stoof, Phys. Rev. Lett. 97, 190407 (2006).
\bibitem{mit3}C. H. Schunck, Y. Shin, A. Schirotzek, M. W. Zwierlein, and W. Ketterle, Science 316, 867 (2007).

\bibitem{review}R. Casalbuoni and G. Nardulli, Rev. Mod. Phys. 76, 263 (2004).
\bibitem{sheehy} D. E. Sheehy and L. Radzihovsky, Ann. Phys. 322, 1790 (2007), and references therein.

\bibitem{orso}G. Orso, Phys. Rev. Lett. 98, 070402 (2007).
\bibitem{hu}H. Hu, X.-J. Liu, and P. D. Drummond, Phys. Rev. Lett. 98, 070403 (2007); Phys. Rev. A 76, 043605 (2007).
\bibitem{guan}X. W. Guan et al, Phys. Rev. B 76, 085120 (2007).

\bibitem{feiguin} A. E. Feiguin and F. Heidrich-Meisner, Phys. Rev. B 76, 220508 (R) (2007);
M. Tezuka and M. Ueda1, Phys. Rev. Lett. 100, 110403 (2008).

\bibitem{fazio} M. Rizzi, M. Polini, M. A. Cazalilla, M. R. Bakhtiari, M.P. Tosi, R. Fazio, Phys. Rev. B 77, 245105 (2008).

\bibitem{parish} M. M. Parish, S. K. Baur, E. J. Mueller, and D. A. Huse, Phys. Rev. Lett. 99, 250403 (2007).

\bibitem{esslinger} H. Moritz, T. St\"oferle, K. G\"unter, M. K\"ohl, and T. Esslinger, Phys. Rev. Lett. 94, 210401 (2005).

\bibitem{schulz}H. J. Schulz and C. Bourbonnais, Phys. Rev. B 27, 5856 (1983);
E. W. Carlson, D. Orgad, S. A. Kivelson, and V. J. Emery, Phys. Rev. B 62, 3422 (2000).

\bibitem{kyang} K. Yang, Phys. Rev. B 63, 140511(R) (2001).
\bibitem{cny}C. N. Yang, Phys. Rev. Lett. 19, 1312 (1967).
\bibitem{gaudin} M. Gaudin, Phys. Lett. A 24, 55 (1967).

\bibitem{olsh}M. Olshanii, Phys. Rev. Lett. 81, 938 (1998).
\bibitem{olsh2}T. Bergeman, M. G. Moore, and M. Olshanii, Phys. Rev. Lett. 91, 163201 (2003).

\bibitem{ceperley} M. Casula, D. M. Ceperley, E. J. Mueller, Phys. Rev. A 78, 033607 (2008).

\bibitem{essler_book} F. H. L. Essler, H. Frahm, F. G\"ohmann, A. Kl\"umper, and V. E. Korepin, The one dimensional Hubbard model, Cambridge University Press, Cambridge (2005).

\bibitem{hulet} R. Hulet, private communications.
\bibitem{giamarchi}T. Giamarchi, Quantum physics in one dimension, Oxford University Press, Oxford (2004).
\bibitem{bible} A. O. Gogolin, A. A. Nersesyan, A. M. Tsvelik, Bosonization and strongly correlated systems, Cambridge University Press, Cambridge (1998).

\bibitem{haldane}F. D. M. Haldane, J. Phys. A 15, 507 (1982).
\bibitem{japa} G. I. Japaridze and A. A. Nersesyan, JETP Lett. 27, 334 (1978).

\bibitem{rainer}H. Burkhardt and D. Rainer, Ann. Phys. 3, 181 (1994).

\bibitem{frahm}H. Frahm and V. E. Korepin, Phys. Rev. B 42, 10 553 (1990); Phys. Rev. B 43, 5653 (1991).
\bibitem{pham}K.-V. Pham, M. Gabay, and P. Lederer, Phys. Rev. B 61, 16397 (2000).

\bibitem{vekua}H. Frahm and T. Vekua, J. Stat. Mech. P01007, (2008).
\bibitem{furusaki}T. Hikihara, A. Furusaki, and K. A. Matveev, Phys. Rev. B 72, 035301 (2005).

\bibitem{taka} M. Takahashi, Thermodynamics of one-dimensional solvable models, Cambridge University Press, Cambridge (1999).

\bibitem{woy2}T. B. Bahder and F. Woynarovich, Phys. Rev. B 33, 2114 (1986).

\bibitem{emery} V. J. Emery, Phys. Rev. B 14, 2989 (1976); A. Moreo and D. J. Scalapino, Phys. Rev. Lett. 98, 216402 (2007);

\bibitem{luscher} A. L\"uscher, R. M. Noack, and A. M. L\"auchli, Phys. Rev. A 78, 013637 (2008).

\bibitem{penc}K. Penc and J. S\'olyom, Phys. Rev. B 47, 6273 (1993).

\bibitem{voit} J. Voit, Y. Wang, and M. Grioni, Phys. Rev. B 61, 7930 (2000).

\bibitem{wen} X.-G. Wen, Phys. Rev. B 42, 6623 (1990).
\bibitem{boies}D. Boies, C. Bourbonnais, and A.-M. S. Tremblay, Phys. Rev. Lett. 74, 968 (1995).


\end{thebibliography}
\end{document}